\documentclass[pra,twocolumn,aps,superscriptaddress,showpacs,epsfig]{revtex4}
\usepackage{epsfig,graphicx}
\usepackage{amsfonts}
\usepackage{amssymb}
\usepackage{amstext}
\usepackage{amsmath}

\begin{document}

\title{Symmetry breaking and self-trapping of a dipolar Bose-Einstein
condensate in a double-well potential}
\author{Bo Xiong}
\affiliation{Department of Mathematics, National University of Singapore, 117543,
Singapore}
\affiliation{Centre of Computational Science and Engineering, National University of
Singapore, 117542, Singapore}
\author{Jiangbin Gong}
\affiliation{Department of Physics, National University of Singapore, 117542, Singapore}
\affiliation{Centre of Computational Science and Engineering, National University of
Singapore, 117542, Singapore}
\affiliation{NUS Graduate School for Integrative Sciences and Engineering, Singapore
117597, Republic of Singapore}
\author{Han Pu}
\affiliation{Department of Physics and Astronomy, and Rice Quantum Institute, Rice
University, Houston, Texas 77251-1892, USA}
\author{Weizhu Bao}
\affiliation{Department of Mathematics, National University of Singapore, 117543,
Singapore}
\affiliation{Centre of Computational Science and Engineering, National University of
Singapore, 117542, Singapore}
\author{Baowen Li}
\affiliation{Department of Physics, National University of Singapore, 117542, Singapore}
\affiliation{Centre of Computational Science and Engineering, National University of
Singapore, 117542, Singapore}
\affiliation{NUS Graduate School for Integrative Sciences and Engineering, Singapore
117597, Republic of Singapore}
\date{\today }

\begin{abstract}
The quantum self-trapping phenomenon of a Bose-Einstein condensate (BEC)
represents a remarkable nonlinear effect of wide interest. By considering a
purely dipolar BEC in a double-well potential, we study how the dipole
orientation affects the ground state structure and the transition between
self-trapping and Josephson oscillation in dynamics. Three-dimensional
numerical results and an effective two-mode model demonstrate that the onset
of self-trapping of a dipolar BEC can be radically modified by the dipole
orientation. We also analyze the failure of the two-mode model in predicting
the rate of Josephson oscillations. We hope that our results can motivate
experimental work as well as future studies of self-trapping of ultracold
dipolar gases in optical lattices.
\end{abstract}

\maketitle

\section{Introduction}

Ultracold dipolar gases display a variety of unique properties that
are absent in those dominated by S-wave scattering
\cite{PuPRL2001more}. The successful Bose-Einstein condensation of
$^{52}$Cr atoms with a large magnetic dipole moment
\cite{GriesmaierPRL2005} and the first experimental realization of
ultracold KRb polar molecular gas \cite{NiScience2008} are
attracting even more experimental and theoretical interests in
ultracold dipolar Bose-Einstein condensates (BECs). Other than
possible applications in fields such as quantum information
\cite{DeMillePRL2002,AndreNP2006}, one main motivation along studies
of ultracold dipolar gases is to explore new physics afforded by the
anisotropic and long-range dipole-dipole interaction. For a recent
review, see Ref.~\cite{reviewMA}. Indeed, by varying the shape of a
dipolar BEC, the dipole polarization axis, and the trapping geometry
\cite{YiPRA2000,GoralPRA2000,SantosPRL2000},
the partially attractive and partially repulsive dipole-dipole interaction
of a dipolar BEC can be easily manipulated to a great extent. Even more
dramatically, by using the well-established Feshbach resonance technique
\cite{WernerPRL2005,LahayeNature2007,LahayePRL2008,KochNP2008}, it is now
possible to significantly reduce or completely shut off the effects of
short-range interactions and hence realize purely dipolar gases, where the
physics is completely dominated by the dipolar interaction. For example,
superfluid transitions in purely dipolar Fermi gases \cite%
{BaranovPRA2002,BaranovPRL2004}, fractional quantum Hall states in purely
dipolar gas trapped in a rotating optical lattice \cite{BaranovPRL2005}, and
collapse features of a purely dipolar BEC \cite{LahayePRL2008} have been
studied.

In this work, we investigate the properties of a dipolar BEC confined in a
double-well potential. We focus on the structure of the ground state wave
function and the dynamical quantum self-trapping (QST) phenomenon which is
among the most dramatic and counter-intuitive effects induced by the
self-interaction of a BEC. The QST phenomenon has been extensively studied
in non-dipolar BEC's \cite{trappingPRL,flach,GatiJPB2007} and is closely
connected with other physical contexts including the Josephson effect in
superconductors \cite{JosephsonPL1962} and the superfluidity of $^{4}$He
\cite{AndersonRMP1966}. In particular, the QST of a non-dipolar BEC in a
double-well potential, which is sometimes called a Bose-Josephson junction,
is now well understood \cite%
{trappingPRL,RaghavanPRA1999,OstrovskayaPRA2000,gongpra2008} and has been
observed experimentally \cite{AlbiezPRL2005}. From a time-independent point
of view, on the mean-field level the self-interaction of a BEC can induce
the emergence of stationary states with a large population imbalance between
two wells. From a time-dependent point of view, the Josephson oscillation
between the two wells can be suppressed by the self-interaction of a BEC,
which can be regarded as a quantum Zeno effect because each atom in the BEC
is being \textquotedblleft measured" by all the other atoms.

Our interest here is how the anisotropic nature of the dipolar interaction
can be exploited to manipulate the dipolar BEC in a double-well potential,
as manifested in the static properties of ground state structure and in the
dynamical evolution of the system prepared out of equilibrium. We approach
this problem by three-dimensional numerical simulations. We find a wide
region where the dynamics of a dipolar BEC can show either Josephson
tunneling or QST without displaying signs of collapse. More significantly,
we shall demonstrate that the dipole orientation can radically affect the
transition from QST to Josephson oscillation. Furthermore, to gain more
insights into the dynamical QST, we construct an effective two-mode model
which has been widely applied in non-dipolar systems. We find that this
simple two-mode model successfully explains the numerically observed
transition threshold between Josephson oscillation and QST, but fails to
account for the oscillation frequency in the Josephson oscillation regime.
We provide a thorough analysis of the success and the failure of the
two-mode model.

This paper is organized as follows. In Sec. II, after introducing a
three-dimensional realistic model of a purely dipolar BEC trapped in a
double-well potential, we study the structure of the ground state wave
function. In Sec. III, we present detailed simulation results regarding how
the dipole orientation affects the self-trapping of the system. In Sec. IV,
we attempt to use a simple effective two-mode model to explain the
dependence of the onset of self-trapping of a dipolar BEC upon the dipole
orientation. In Sec. V we discuss why our effective two-mode model cannot be
used to estimate the oscillation frequency in the Josephson oscillation
regime. Section VI concludes and summarizes this work. Appendix A presents
some details regarding our numerical calculations of the dipole-dipole
interaction potential.

\section{Ground state structure}

To be specific, let us consider a dipolar BEC of $^{52}$Cr, which has large
magnetic moment $\mu =6\mu _{B}$ ($\mu _{B}$ is the Bohr magneton). We
assume below the S-wave scattering length is tuned to zero via Feshbach
resonance, thus resulting in a purely dipolar atomic BEC. We further assume
that the system is confined in a trapping potential with a cylindrical
symmetry in the $x$-$y$ plane and a double-well structure along the $z$
direction. That is, we consider a confining potential
\begin{equation*}
V(x,y,z)=\frac{m}{2}(\omega _{\rho }^{2}x^{2}+\omega _{\rho
}^{2}y^{2}+\omega _{z}^{2}z^{2})+A\exp (-z^{2}/2\sigma _{0}^{2}),
\end{equation*}%
where $m$ is the atomic mass, $\omega _{x}=\omega _{y}=\omega _{\rho }$ is
the confining harmonic frequency in the transverse direction, $\omega _{z}$
is the confining harmonic frequency in the longitudinal direction, $A$ and $%
\sigma _{0}$ are the height and the width of a Gaussian profile modeling a
barrier between the two potential wells along the $z$ direction. Without
loss of generality, we assume that the dipole moments are polarized by the
external magnetic field and are confined in the $x$-$z$ plane. The two-body
dipolar interaction potential is then given by
\begin{equation}
U_{dd}(\mathbf{r})=d^{2}\left[ r^{2}-3(z\cos \varphi +x\sin \varphi )^{2}%
\right] /r^{5},  \label{DDI}
\end{equation}%
where $\varphi $ is the angle between the polarized dipole orientation and
the $z$ axis, $d^{2}=\mu _{0}\mu ^{2}/4\pi $ with $\mu _{0}$ being the
magnetic permeability of the vacuum. For convenience we define a
dimensionless dipolar interaction parameter, $D=(N-1)md^{2}/(\hbar
^{2}a_{ho})$, where $a_{ho}=\sqrt{\hbar /(m\omega _{z})}$ is the axial
harmonic oscillator length and $N$ is the total number of atoms. We can then
adopt a unit system where the units for length, time, and energy are given
by $a_{ho}$, $1/\omega _{z}$ and $\hbar \omega _{z}$, respectively. Unless
specified otherwise, for the numerical results presented below we set $%
\omega _{\rho }/\omega _{z}=10$, $A=4\hbar \omega _{z}$, $\sigma
_{0}=0.2a_{ho}$, and $D=0.6$ (which corresponds to about 2000 Cr atoms for
the axial trapping frequency $\omega _{z}=2\pi \times 330$ Hz). In such a
confining potential, the mean-field dynamics of a purely dipolar BEC at $T=0$
is described by the following dimensionless time-dependent Gross-Pitaevskii
equation:
\begin{eqnarray}
i\frac{\partial \Psi (\mathbf{r},t)}{\partial t} &=&-\frac{1}{2}\nabla
^{2}\Psi (\mathbf{r},t)+V(\mathbf{r})\Psi (\mathbf{r},t)  \notag \\
&&+\int U_{dd}(\mathbf{r}-\mathbf{r}^{\prime })|\Psi (\mathbf{r}^{\prime
},t)|^{2}\Psi (\mathbf{r},t)d^{3}\mathbf{r}^{\prime },  \label{eq:GPE}
\end{eqnarray}%
where the macroscopic wave function $\Psi $ is normalized to unity.

\begin{figure}[tbp]
\epsfxsize=10.0cm \centerline{\epsffile{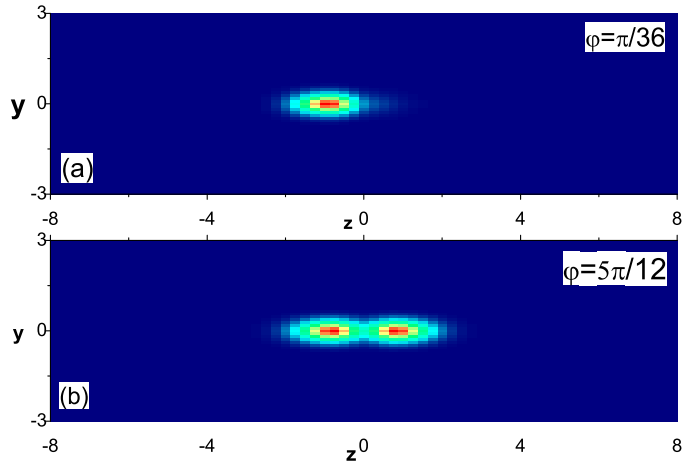}}
\caption{(Color online) Column density of the ground state density profile $%
\protect\int |\Psi (\mathbf{r})|^2 \,dx$ for two different angles $\protect%
\varphi$.} \label{gswf}
\end{figure}

\begin{figure}[tbp]
\epsfxsize=9.0cm \centerline{\epsffile{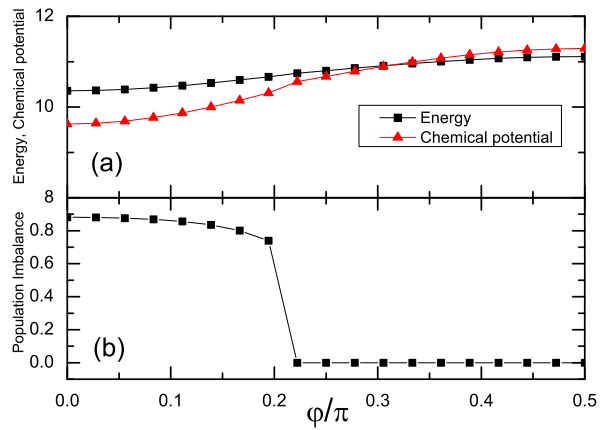}}
\caption{(Color online) The ground state energy, chemical potential
as well
as the population difference between the two wells as functions of $\protect%
\varphi$.} \label{gs}
\end{figure}

In our numerical simulations we adopt the time-splitting Fourier
pseudospectral method \cite{FornbergBook1996, BaoJCP2003}\textbf{.}
Discretization used in our calculations is $\Delta t=0.005$ for time, and $%
\Delta x=\Delta y=0.09375$, $\Delta z=0.25$ in space. The calculation is
performed in a box of size ($L_{x}$, $L_{y}$, $L_{z}$). The values of $L_{x}$%
, $L_{y}$, and $L_{z}$ are chosen such that the box is much larger
than the size of the trapped condensate. We impose zero boundary
conditions for the wavefunction amplitudes at $x=\pm
L_{x}/2$, $y=\pm L_{y}/2$, and $z=\pm L_{z}/2$. Typically we set $%
L_{x}=L_{y}=6$, $L_{z}=16$. To evaluate the dipolar interaction potential
that involves a convolution integral, we use the following fast Fourier
transform technique,%
\begin{equation}
\int U_{dd}(\mathbf{r}-\mathbf{r}^{\prime })|\Psi (\mathbf{r}^{\prime
},t)|^{2}d^{3}\mathbf{r}^{\prime }=F^{-1}\{F[U_{dd}(\mathbf{r})]F[|\Psi (%
\mathbf{r},t)|^{2}]\},  \label{DDICONV}
\end{equation}%
where $F$ and $F^{-1}$ stand for fast Fourier transform and fast inverse
Fourier transform, respectively. The term $F[U_{dd}(\mathbf{r})]$ in Eq. (%
\ref{DDICONV}) is calculated analytically in the momentum space (see detail
in Appendix):
\begin{eqnarray}
F[U_{dd}(\mathbf{r})] &=&\int U_{dd}\left( \mathbf{r}\right) e^{i\mathbf{%
k\cdot r}}d^{3}\mathbf{r}  \notag \\
&=&\pi d^{2}\left[ 2\sin ^{2}\varphi \sin ^{2}\theta _{\mathbf{k}}\cos
\left( 2\phi _{\mathbf{k}}\right) \right.  \notag \\
&+&2\sin (2\varphi )\sin (2\theta _{\mathbf{k}})\cos \phi
_{\mathbf{k}}
\notag \\
&+&\left. \left( 4/3-2\sin ^{2}\varphi \right) \left( 3\cos ^{2}\theta _{%
\mathbf{k}}-1\right) \right] ,  \label{FDDICONV}
\end{eqnarray}%
where $\theta _{\mathbf{k}}$ and $\phi _{\mathbf{k}}$ are the polar angle
and the azimuthal angle in the spherical coordinate system for the momentum
space.

The ground state is obtained by evolving Eq.~(\ref{eq:GPE}) in
imaginary time. Figure \ref{gswf} represents the ground state wave
functions for two different dipolar angles: $\varphi =\pi /36$ and
$\varphi =5\pi /12$. One can clearly see that in the former case,
the wave function is asymmetric and localized in one of the wells;
while for the latter, the wave function is symmetric and the
population is equally distributed in both wells. Given the geometry
of the trapping potential, the effective dipolar interaction is
predominantly attractive for small values of $\varphi $ and
repulsive for large values of $\varphi $. This explains the
different structures of the ground state wave function at different
angles. Figure \ref{gs} shows the ground state energy, chemical
potential as well as the population difference between the two wells
as functions of $\varphi $. It indicates that the
symmetric/asymmetric transition as induced by the variation of
$\varphi $ is a continuous one and the critical angle is about
$\varphi =0.22\pi $.

\begin{figure}[tbp]
\epsfxsize=13.cm \centerline{\epsffile{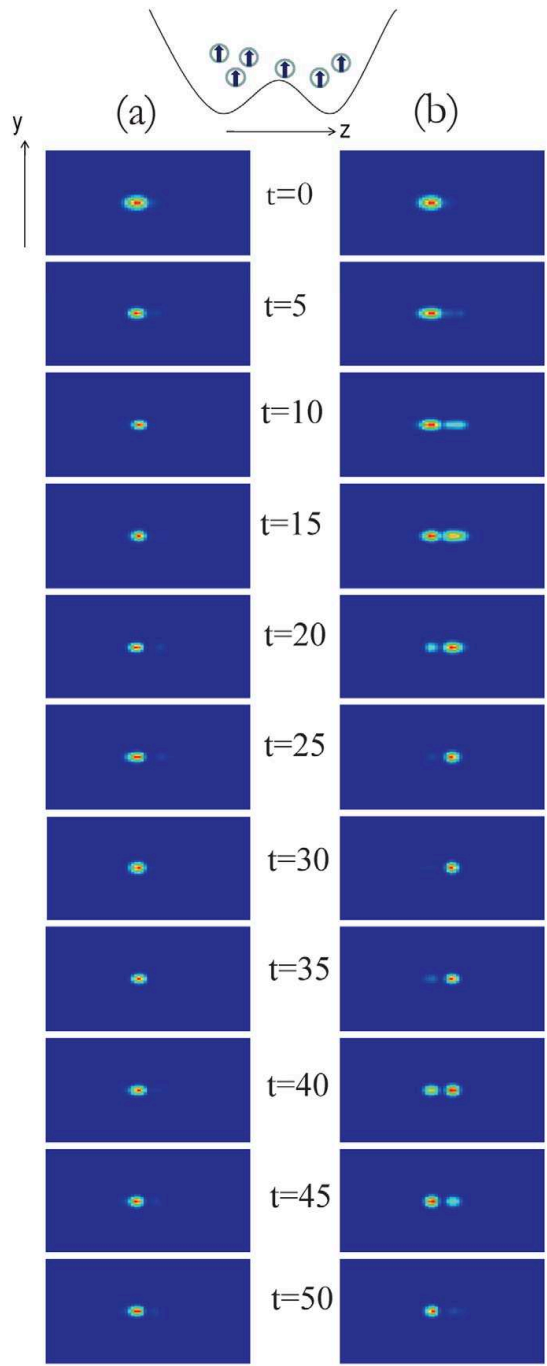}}
\caption{(Color online) Column density of the condensate. (a) For
the dipole orientation angle $\protect\varphi =\protect\pi /4$, the
population is mainly trapped in the left well instead of tunneling
between the two wells.
(b) Changing only the dipole orientation parameter to $\protect\varphi =5%
\protect\pi /12$, the tunneling between the two wells, or the Josephson
oscillation, is observed. See the text for other system parameters.}
\label{Fig. 1}
\end{figure}

\section{Self-trapping and Josephson Oscillations}

We now turn to the dynamical properties of the system. We simulate the
following situation. Initially we prepare the system in the ground state
with $\varphi=\pi/36$. As shown in Fig.~\ref{gswf}(a), for this dipole
angle, the wave function is localized in one of the wells (for the example
given, the atoms are localized in the left well). At $t=0$, we suddenly
change $\varphi$ to some other value and study the ensuing dynamics of the
system.

In Fig. \ref{Fig. 1}, we show the time evolution of $\Psi (\mathbf{r})$ for
two different final values of the dipole orientation parameter $\varphi $.
In Fig. \ref{Fig. 1}(a) $\varphi $ is chosen to be $\pi /4$, and the
condensate is found to remain localized in the left well for all times even
though the ground state for $\varphi=\pi/4$ should have population equally
distributed in both wells [see Fig.~\ref{gs}(b)]. The system is hence
clearly in the QST regime. By contrast, in the case of Fig. \ref{Fig. 1}(b),
$\varphi $ is changed to $5\pi /12$ and the population exhibits regular
oscillations. The system in this case is clearly in the Josephson
oscillation regime. Because the only parameter difference between Fig. \ref%
{Fig. 1}(a) and Fig. \ref{Fig. 1}(b) is the final value of $\varphi $, Fig. %
\ref{Fig. 1} vividly demonstrates the tunability of the system through the
orientation of the dipoles, a feature obviously absent in non-dipolar BEC's.

\begin{figure}[tbp]
\epsfxsize=11cm \centerline{\epsffile{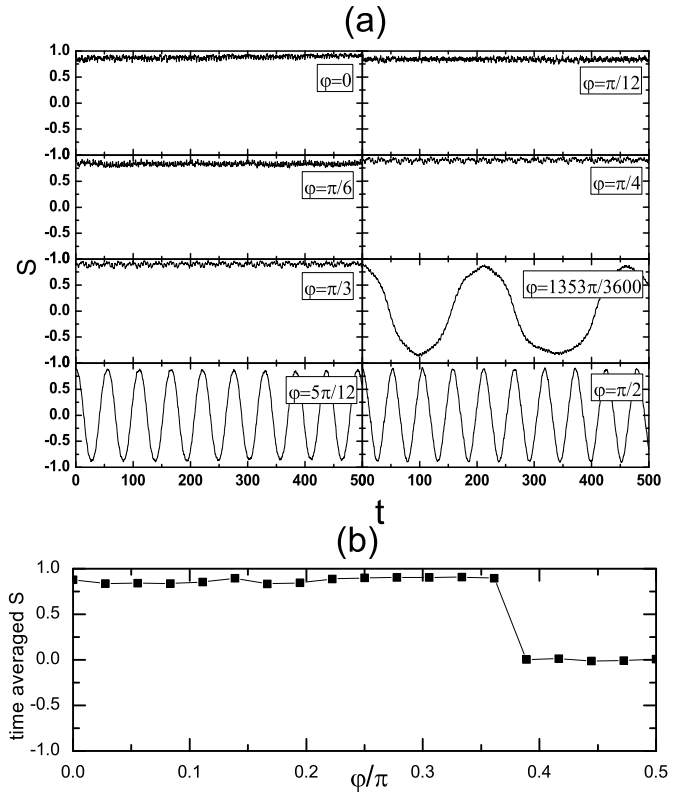}} \caption{(a)
Population imbalance $S(t)$ versus time, for different values of the
dipole orientation parameter $\protect\varphi $. As $\protect\varphi
$ exceeds a critical value $\protect\varphi_c \approx 0.37
\protect\pi$, the
population imbalance starts to oscillate around zero. (b) Time averaged $%
S(t) $ as a function of the dipole orientation parameter $\protect\varphi $.
When $\protect\varphi $ exceeds $\protect\varphi_c$, the time averaged
population imbalance suddenly decreases to zero. The transition between
Josephson oscillation and quantum self-trapping can hence be manipulated by
tuning the dipole orientation. Other system parameters are the same as those
used in Fig. 1.}
\label{Fig. 2}
\end{figure}

To examine in more detail how the dipole orientation impacts on the
QST, we next scan the value of $\varphi $ and then plot the
corresponding time evolution of a normalized population imbalance
$S$ between the two wells, namely, the population difference divided
by the total number of atoms. Results are shown in Fig. \ref{Fig.
2}. It can be seen that there exists a critical angle $\varphi
_{c}\approx 0.37\pi $: for $\varphi <\varphi _{c}$, the system is in
the QST regime; for $\varphi >\varphi _{c}$, it enters the Josephson
oscillation regime. We have also checked the critical regime in more
detail by scanning $\varphi $ in smaller steps. In particular, we
also show in Fig. \ref{Fig. 2}(a) the population dynamics for
$\varphi =1353\pi /3600$, a value just above $\varphi _{c}$.
Interestingly, the oscillation dynamics for that case shows
noteworthy  critical oscillation behavior. Evidently then, our
detailed results here further confirm that the dipole orientation of
a dipolar BEC offers a simple and powerful means to manipulate the
transition from QST to the Josephson oscillation.

We have also studied many cases with other trapping frequency
ratios. For example, we let $\omega _{\rho }/\omega _{z}$ vary from
$10.0$ to $5.0$ via increasing  $\omega _{z}$. It is found that the
frequency of the Josephson oscillation can be very sensitive to the
trapping frequency ratio. For example, for $\omega _{\rho }/\omega
_{z}=5.882$, $\varphi =5\pi /12$, and the width of the initial
wavefunction in $z$ narrowed by $\sqrt{10/5.882}$, the Josephson
oscillation frequency is more than 1.7 times larger than that shown
in Fig. \ref{Fig. 2} (a) for the same value of $\varphi$. If $\omega
_{\rho }/\omega _{z}=5.882$ and if the width of the initial
wavefunction in $z$ is not narrowed , then with all other parameters
fixed, the condensate can already show clear signs of collapse
(i.e., developing very high densities in our mean-field
calculations) as we vary the dipole orientation. Interestingly,
the condensate may collapse before we reach the critical point $%
\varphi _{c}$ for the transition between QST and the Josephson
oscillation. This being the case, a stable dipolar BEC in a
double-well potential may be always in the QST regime.

\section{A Simple Effective Two-Mode Model}

To understand and gain more insights into the dynamical behavior presented
in the previous section, we now construct a two-mode model that has been
widely used for the study of QST for non-dipolar BEC's. To that end we
assume that the main feature of the time-evolving wave-function $\Psi (%
\mathbf{r},t)$ is captured by two normalized real basis states,
\begin{equation}
\Psi (\mathbf{r},t)=\psi _{1}(t)\Phi _{1}(\mathbf{r})+\psi _{2}(t)\Phi _{2}(%
\mathbf{r}),  \label{eq:psi12defn}
\end{equation}%
where $\psi _{1,2}(t)=\sqrt{N_{1,2}(t)}e^{i\theta _{1,2}(t)}$, and $\Phi
_{1,2}(r)$ are localized in each of the two wells. The total number of atoms
is given by $|\psi _{1}|^{2}+|\psi _{2}|^{2}=N_{T}$. Substituting Eq. (\ref%
{eq:psi12defn}) into Eq. (\ref{eq:GPE}), we obtain
\begin{eqnarray}
&&i\frac{d\psi _{1}(t)}{dt}{\Phi _{1}(\mathbf{r})}+i\frac{d\psi _{2}(t)}{dt}{%
\Phi _{2}(\mathbf{r})}\hspace*{1.2cm}  \notag \\
&=&-{\frac{1}{{2}}}\left[ \psi _{1}(t)\nabla ^{2}{\Phi _{1}(\mathbf{r})+}%
\psi _{2}(t)\nabla ^{2}{\Phi _{2}(\mathbf{r})}\right]   \notag \\
&&+\left[ {\psi _{1}(t)V(\mathbf{r})\Phi _{1}(\mathbf{r})+\psi _{2}(t)V(%
\mathbf{r})\Phi _{2}(\mathbf{r})}\right]   \notag \\
&+&\left\{ |\psi _{1}|^{2}\psi _{1}{\Phi _{1}(\mathbf{r})}\int U_{dd}(%
\mathbf{r}-\mathbf{r}^{\prime })\left\vert {\Phi _{1}(}\mathbf{r}^{\prime }{)%
}\right\vert ^{2}d^{3}\mathbf{r}^{\prime }\right.   \notag \\
&+&|\psi _{1}|^{2}\psi _{2}[{\Phi _{2}(}\mathbf{r})\int U_{dd}(\mathbf{r}-%
\mathbf{r}^{\prime })\left\vert {\Phi _{1}(}\mathbf{r}^{\prime }{)}%
\right\vert ^{2}d^{3}\mathbf{r}^{\prime }  \notag \\
&+&{\Phi _{1}(\mathbf{r})}\int U_{dd}(\mathbf{r}-\mathbf{r}^{\prime }){\Phi
_{1}(}\mathbf{r}^{\prime }{)\Phi _{2}(}\mathbf{r}^{\prime })d^{3}\mathbf{r}%
^{\prime }]  \notag \\
&&+\psi _{1}^{2}{}\psi _{2}{\Phi _{1}(}\mathbf{r})\int U_{dd}(\mathbf{r}-%
\mathbf{r}^{\prime }){\Phi _{1}(}\mathbf{r}^{\prime }){\Phi _{2}(}\mathbf{r}%
^{\prime }{)}d^{3}\mathbf{r}^{\prime }  \notag \\
&+&\psi _{1}|\psi _{2}|^{2}[{\Phi _{2}(}\mathbf{r})\int U_{dd}(\mathbf{r}-%
\mathbf{r}^{\prime }){\Phi _{1}(}\mathbf{r}^{\prime }){\Phi _{2}(}\mathbf{r}%
^{\prime }{)}d^{3}\mathbf{r}^{\prime }  \notag \\
&+&{\Phi _{1}(\mathbf{r})}\int U_{dd}(\mathbf{r}-\mathbf{r}^{\prime
})\left\vert {\Phi _{2}(}\mathbf{r}^{\prime }{)}\right\vert ^{2}d^{3}\mathbf{%
r}^{\prime }]  \notag \\
&+&\psi _{1}^{\star }\psi _{2}^{2}{\Phi _{2}(}\mathbf{r})\int U_{dd}(\mathbf{%
r}-\mathbf{r}^{\prime }){\Phi _{1}(}\mathbf{r}^{\prime }){\Phi _{2}(}\mathbf{%
r}^{\prime }{)}d^{3}\mathbf{r}^{\prime }  \notag \\
&+&\left. |\psi _{2}|^{2}\psi _{2}{\Phi _{2}(\mathbf{r})}\int U_{dd}(\mathbf{%
r}-\mathbf{r}^{\prime })\left\vert {\Phi _{2}(}\mathbf{r}^{\prime }{)}%
\right\vert ^{2}d^{3}\mathbf{r}^{\prime }\right\} .  \label{BJJ-1}
\end{eqnarray}%
As in previous studies of QST of non-dipolar BEC's, if we assume the overlap
between the two modes being zero, namely,
\begin{eqnarray}
&&\int \Phi _{i}{(\mathbf{r})}\Phi _{j}{(\mathbf{r})}d^{3}\mathbf{r}\approx
\delta _{i,j};\text{ }i,j=1,2,  \notag \\
&&\int f(\mathbf{r})\Phi _{i}{(\mathbf{r})}\Phi _{j}{(\mathbf{r})}d^{3}%
\mathbf{r}\approx 0;\text{ }i\neq j;  \label{eq:va3}
\end{eqnarray}%
then most of the terms in Eq. (\ref{BJJ-1}) will vanish.
Indeed, such a great simplification is the main advantage of a two-mode
picture in the first place. Adopting this zero-overlap assumption, we obtain
two simple coupled equations for $\psi _{1,2}(t)$:
\begin{subequations}
\label{eq:BJJ}
\begin{eqnarray}
i\frac{\partial \psi _{1}}{\partial t}
&=&[(E_{1}^{0}+B_{12})+(A_{11}-B_{12})|\psi _{1}|^{2}]\psi _{1}-\kappa \psi
_{2},  \notag \\
&&  \label{eq:BJJa} \\
i\frac{\partial \psi _{2}}{\partial t}
&=&[(E_{2}^{0}+B_{21})+(A_{22}-B_{21})|\psi _{2}|^{2})]\psi _{2}-\kappa \psi
_{1}.  \notag \\
&&  \label{eq:BJJb}
\end{eqnarray}%
The parameters in the above two-mode equations of motion are given by
\end{subequations}
\begin{eqnarray}
E_{i}^{0} &=&\int {{\frac{1}{{2}}}\left[ (\nabla \Phi _{i})^{2}+|\Phi _{i}(%
\mathbf{r})|^{2}V(\mathbf{r})\right] d^{3}\mathbf{r},}  \label{BJJ-E} \\
A_{ii} &=&\int U_{dd}(\mathbf{r}-\mathbf{r}^{\prime })|{\Phi _{i}(\mathbf{r})%
}|^{2}|{\Phi _{i}(\mathbf{r}}^{\prime }{)}|^{2}d^{3}\mathbf{r}d^{3}\mathbf{r}%
^{\prime },  \label{BJJ-A} \\
B_{12} &=&B_{21}=\int U_{dd}(\mathbf{r}-\mathbf{r}^{\prime })|{\Phi _{1}(%
\mathbf{r})}|^{2}|{\Phi _{2}(\mathbf{r}}^{\prime }{)}|^{2}d^{3}\mathbf{r}%
d^{3}\mathbf{r}^{\prime },  \notag \\
&&  \label{BJJ-B} \\
\kappa  &\simeq &-\int {\ \left[ {\frac{1}{{2}}}(\nabla {\Phi }_{1}\nabla {%
\Phi }_{2})+{\Phi }_{1}V(\mathbf{r}){\Phi }_{2}\right] d^{3}\mathbf{r}}\,.
\label{BJJC}
\end{eqnarray}%
Here, $A_{ii}$ and $B_{12}$ are the \textquotedblleft on-site" and
\textquotedblleft off-site" contribution of the dipolar interaction term,
respectively. The presence of the off-site interaction term $B_{12}$ is a
unique feature for long-range interactions. $E_{i}^{0}+B_{12}$ can be
regarded as an effective on-site energy for the two wells, and $A_{ii}-B_{12}
$ can be regarded as an effective self-interaction strength, and $\kappa $
is the coupling strength between the two modes and hence gives the
characteristic rate of oscillations between the two modes if the nonlinear
terms are neglected. Note that the expressions in Eq.~(\ref{eq:BJJ}) are
very similar to those for a non-dipolar BEC in a double-well potential \cite%
{trappingPRL} except that the off-site interaction term $B_{12}$ is absent
in non-dipolar systems.

In terms of the population imbalance $S(t)=[|\psi _{1}(t)|^{2}-|\psi
_{2}(t)|^{2}]/N_{T}$ and a relative phase parameter $\phi (t)\equiv \theta
_{2}(t)-\theta _{1}(t)$, Eqs. (\ref{eq:BJJa}) and (\ref{eq:BJJb}) assume an
enlightening form,
\begin{subequations}
\label{eq:zphidot}
\begin{eqnarray}
\dot{S}(t) &=&-\sqrt{1-S^{2}(t)}\sin [\phi (t)],  \label{eq:zphidota} \\
\dot{\phi}(t) &=&\Delta E+\Lambda S(t)+\frac{S(t)}{\sqrt{1-S^{2}(t)}}\cos
[\phi (t)],  \label{eq:zphidotb}
\end{eqnarray}%
where the time variable is rescaled by a factor $2\kappa $, with parameters $%
\Delta E$ and $\Lambda $ defined by
\end{subequations}
\begin{subequations}
\label{eq:delElmbdef}
\begin{eqnarray}
\Delta E &=&\frac{E_{1}^{0}-E_{2}^{0}}{2\kappa }+\frac{(A_{11}-A_{22})N_{T}}{%
4\kappa },  \label{TM_DE} \\
\Lambda &=&\frac{(A_{11}+A_{22}-2B_{12})N_{T}}{4\kappa }.  \label{TM_namda}
\end{eqnarray}%
Because Eqs. (\ref{eq:zphidota}) and (\ref{eq:zphidotb}) are exactly the
same as those for a two-mode QST model of a non-dipolar BEC, it is clear
that within the above zero-overlap approximation, the underlying physics of
the QST dynamics of a dipolar BEC should be similar to that of a non-dipolar
BEC. Indeed, the mechanical analog of Eqs. (\ref{eq:zphidota}) and (\ref%
{eq:zphidotb}) is a classical non-rigid pendulum of tilt angle $\phi
$,
angular momentum $S$, and a length proportional to $\sqrt{1-S^{2}(t)}$ \cite%
{trappingPRL}. For an initial condition $[S(0),\phi (0)]$, this
pendulum will oscillate around $S=0$ if and only if $\Lambda
<\Lambda _{c}$, where
\end{subequations}
\begin{equation}
\Lambda _{c}=\frac{2[1-\Delta ES(0)+\sqrt{1-S(0)^{2}}\cos \phi (0)]}{S(0)^{2}%
}\,.  \label{eq:lmbself-free}
\end{equation}%
Translating back to our BEC context, we have that if $\Lambda
>\Lambda _{c}$, then the system will be in the QST regime; and if
$\Lambda <\Lambda _{c}$, then the Josephson oscillations can be
expected.

The above two-mode analysis also indicates that the QST phenomenon of a
dipolar BEC can be easily manipulated, because the important parameter $%
\Lambda $ depends strongly on the dipole orientation as well as the trap
geometry. Furthermore, the expression of $\Lambda $ in Eq. (\ref{TM_namda})
implies that the competition between the on-site interaction terms $A_{ii}$
and the off-site long-range term $B_{12}$ might play a role in the QST
physics. For example, if under some circumstances the on-site contribution
cancels out the off-site contribution, then $\Lambda $ will be small and
hence QST will not occur.


Our discussions so far are based on the strong assumption that a two-mode
model still applies well to a purely dipolar BEC in a three-dimensional
potential with a double-well structure. However, because our simulations
show that the density profile in each well depends strongly on the
population imbalance, a simple two-mode model is not expected to work
satisfactorily. As such, we propose an \textit{effective} two-mode mode
constructed in a self-consistent manner, by first extracting useful
information from our numerical simulations. Specifically, we use the
following normalized Gaussian ansatz
\begin{equation}
{\Phi _{i}(\mathbf{r})}=\frac{%
e^{[-(x+x_{i})^{2}/2a_{i}^{2}-(y+y_{i})^{2}/2b_{i}^{2}-(z+z_{i})^{2}/2c_{i}^{2}]}%
}{\sqrt{a_{i}b_{i}c_{i}}\pi ^{\frac{3}{4}}},  \label{GauAn}
\end{equation}%
to model the basis states $\Phi _{1,2}(\mathbf{r})$ in the two-mode model,
with its parameters $a_{i},b_{i},c_{i},x_{i},y_{i}$, and $z_{i}$ to be
fitted by long-time average properties of the density profile in our full
three-dimensional simulations of Eq. (\ref{eq:GPE}). Once the parameters for
the two basis states are obtained numerically, then the value of $\kappa $
can be obtained analytically, and the values of $A_{ii}$ and $B_{12}$, and
hence the value of $\Lambda $, can all be obtained. In so doing, the
location and width of the Gaussian ansatz for ${\Phi _{i}(\mathbf{r})}$ may
change with the initial condition, e.g., the initial population imbalance.
This feature is hence outside the conventional two-mode model of QST. For
that reason our self-consistent effective two-mode model is expected to
catch some features not available in a conventional two-mode model.

\begin{figure}[tbp]
\epsfxsize=9.5cm \centerline{\epsffile{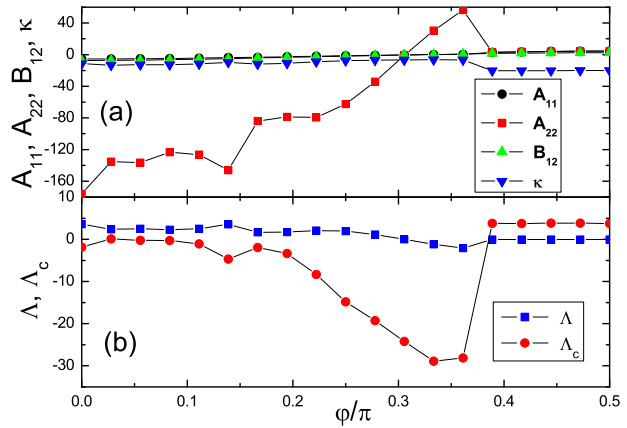}}
\caption{(Color online) (a) Values for the two-mode model parameters $A_{11}$%
, $A_{22}$, $B_{12}$, $\protect\kappa $, as functions of the dipolar
orientation parameter $\protect\varphi $, calculated by first fitting the
Gaussian ansatz in (\protect\ref{GauAn}) with the time-averaged properties
of the density profile in our three-dimensional simulations. (b) Value of $%
\Lambda $ calculated from Eq. (\protect\ref{TM_namda}) and value of $\Lambda
_{c}$ calculated from Eq. (\protect\ref{eq:lmbself-free}) with $\protect\phi %
(0)=0$, as a function of the dipolar orientation parameter $\protect\varphi $%
. Note that for $\protect\varphi <\protect\varphi _{c}\approx 0.39\protect%
\pi $, $\Lambda >\Lambda _{c}$; and for $\protect\varphi >\protect\varphi %
_{c}$, $\Lambda <\Lambda _{c}$.}
\label{Fig. 3}
\end{figure}

The results from our effective two-mode model are presented in Fig. \ref%
{Fig. 3}. Figure \ref{Fig. 3}(a) depicts how the values of the two-mode
model parameters $A_{11}$, $A_{22}$, $B_{12}$ and $\kappa $ change with the
dipole orientation parameter $\varphi $. Figure \ref{Fig. 3}(b) shows the $%
\varphi $-dependence of $\Lambda $ as well as $\Lambda _{c}$. From Fig. \ref%
{Fig. 3}(b) it is seen that the value of $\Lambda $ initially
increases, and then decreases to almost zero at about $\varphi
\approx 0.37\pi$. On
the other hand, the value of $\Lambda _{c}$, calculated from Eq.~(\ref%
{eq:lmbself-free}) with $\phi (0)=0$ (as an example), is seen to be
less than $\Lambda $ initially, and then exceeds $\Lambda $ at about
$\varphi \approx 0.39\pi$. Therefore, as we tune the dipole
orientation, our
effective two-mode model displays a switch from $\Lambda >\Lambda _{c}$ to $%
\Lambda <\Lambda _{c}$, thereby predicting the transition from the
self-trapping regime to the Josephson oscillation regime at a
critical value $\varphi \approx 0.39\pi$. The critical value
obtained from the two-mode model matches with our previous numerical
simulations of Eq.~(\ref{eq:GPE}), where we observed $\varphi
_{c}\approx 0.37\pi $. We conclude that at least in our effective
two-mode model, the anisotropic nature of QST for a dipolar gas can
be understood in terms of the $\varphi $-dependence of the two-mode
model parameters.

Results here indicate that when predicting the transition point between the
QST regime and the Josephson oscillation regime, the above-mentioned
zero-overlap approximation is acceptable. But it should be noted that this
zero-overlap approximation is a very rough one, and there should be a high
price for that. Indeed, as recently shown via a careful analysis of the
validity of two-mode approximations for one-dimensional systems \cite%
{beyond-two-mode}, the Josephson oscillation rate based on a simple two-mode
theory may differ greatly from the actual rate. With that in mind we ask if
our effective two-mode picture can correctly describe other dynamical
aspects. In particular, can our effective two-mode picture be employed to
predict a right order of magnitude of the frequency of the regular Josephson
oscillations seen in the four bottom panels in Fig. \ref{Fig. 2}(a)? As it
turns out, the answer is negative. For example, for $\varphi =\pi /2$, the
oscillation period observed from Fig. \ref{Fig. 2}(a) is about $t=50$;
whereas the oscillation period from our effective two-mode model is about $%
\pi /\kappa \sim 0.157$ [see Eqs. (\ref{eq:BJJa}) and
(\ref{eq:BJJb})]. The difference is hence more than two orders of
magnitude! Our next section is devoted to this interesting
observation.

\section{Discussions}

Here we provide a detailed analysis of the huge discrepancy between the
two-mode parameter $\kappa$ and the actual rate of the observed Josephson
oscillations. Let us first re-examine the meaning of $\kappa$ defined in Eq.
(\ref{BJJC}). Clearly, for a three-dimensional problem, if the transverse
direction is highly confined, the transverse motion will have a high kinetic
energy and hence a large magnitude of wavefunction gradients with respect to
$x$ and $y$. According to Eq.~(\ref{BJJC}), this large transverse gradient
will directly yield a large magnitude of $\kappa$. Hence, for tightly
confined systems the magnitude of $\kappa$ is largely contributed by the
kinetic energy in the transverse direction. In our two-mode treatment, $%
\kappa$ is identified as the parameter that determines the oscillation rate
of the population between the two wells. As we show below, that $\kappa$
differs significantly from the actual oscillation rate obtained numerically
originates from the zero-overlap approximation made in the two-mode model.

Specifically, if we now drop the previous zero-overlap assumption, Eq. (\ref%
{BJJ-1}) leads to
\begin{subequations}
\label{ITM}
\begin{eqnarray}
i\frac{\partial \psi _{1}}{\partial t}=-\frac{\kappa C_{12}+E_{1}^{0}+B_{12}%
}{C_{12}^{2}-1}\psi _{1} &+&\frac{(E_{2}^{0}+B_{21})C_{12}+\kappa }{%
C_{12}^{2}-1}\psi _{2}  \notag \\
&+&Q_{1}^{\text{nl}},  \label{ITM(a)} \\
i\frac{\partial \psi _{2}}{\partial t}=-\frac{\kappa C_{12}+E_{2}^{0}+B_{21}%
}{C_{12}^{2}-1}\psi _{2} &+&\frac{(E_{1}^{0}+B_{12})C_{12}+\kappa }{%
C_{12}^{2}-1}\psi _{1}  \notag \\
&+&Q_{2}^{\text{nl}},  \label{ITM(b)}
\end{eqnarray}%
where $Q_{1}^{\text{nl}}$ and $Q_{2}^{\text{nl}}$ are the nonlinear terms
given by
\end{subequations}
\begin{eqnarray}
Q_{1}^{\text{nl}} &\equiv &\frac{C_{12}}{C_{12}^{2}-1}\sigma _{2}-\frac{1}{%
C_{12}^{2}-1}\sigma _{1},  \notag \\
Q_{2}^{\text{nl}} &\equiv &\frac{C_{12}}{C_{12}^{2}-1}\sigma _{1}-\frac{1}{%
C_{12}^{2}-1}\sigma _{2}
\end{eqnarray}%
with
\begin{subequations}
\label{Sigma}
\begin{eqnarray}
\sigma _{1} &=&(A_{11}-B_{12})|\psi _{1}|^{2}\psi _{1}+2D_{12}\psi
_{1}^{2}\psi _{2}  \notag \\
&+&D_{12}|\psi _{1}|^{2}\psi _{2}^{\star }+G_{12}|\psi _{2}|^{2}\psi _{1}
\notag \\
&+&G_{12}\psi _{1}^{\star }\psi _{2}^{2}+M_{12}|\psi _{2}|^{2}\psi _{2},
\label{Sigma(a)} \\
\sigma _{2} &=&(A_{22}-B_{12})|\psi _{2}|^{2}\psi _{2}+2D_{21}\psi
_{2}^{2}\psi _{1}  \notag \\
&+&D_{21}\psi _{2}^{2}\psi _{1}^{\star }+G_{21}|\psi _{1}|^{2}\psi _{2}
\notag \\
&+&G_{21}\psi _{2}^{\star }\psi _{1}^{2}+M_{21}|\psi _{1}|^{2}\psi _{1},
\label{Sigma(b)}
\end{eqnarray}%
and
\end{subequations}
\begin{eqnarray*}
C_{12} &=&\int {\Phi _{1}(\mathbf{r})\Phi {_{2}}(\mathbf{r})d^{3}\mathbf{r}}%
\text{,} \\
D_{ij} &=&\int U_{dd}(\mathbf{r}-\mathbf{r}^{\prime }){\Phi _{i}(\mathbf{r}%
)\Phi _{j}(}\mathbf{r})|{\Phi _{i}(}\mathbf{r}^{\prime }{)}|^{2}d^{3}\mathbf{%
r}d^{3}\mathbf{r}^{\prime }, \\
G_{ij} &=&\int U_{dd}(\mathbf{r}-\mathbf{r}^{\prime }){\Phi _{i}(\mathbf{r}%
)\Phi _{j}(}\mathbf{r}){{\Phi _{i}(}\mathbf{r}^{\prime })\Phi _{j}(}\mathbf{r%
}^{\prime }{)d^{3}\mathbf{r}d^{3}\mathbf{r}^{\prime }}\text{,} \\
M_{ij} &=&\int U_{dd}(\mathbf{r}-\mathbf{r}^{\prime }){\Phi _{i}(\mathbf{r}%
)\Phi _{j}(\mathbf{r})|\Phi _{j}(}\mathbf{r}^{\prime }{)}|^{2}d^{3}\mathbf{r}%
d^{3}\mathbf{r}^{\prime }\,.
\end{eqnarray*}

Evidently then, once we drop the zero-overlap assumption, the two-mode
treatment can no longer give a simple picture of the dynamics. Nevertheless,
Eqs. (\ref{ITM(a)}) and (\ref{ITM(b)}) indicate that if we take into account
the nonzero overlap $C_{12}$, the effective coupling between the two modes
will be quite different. Indeed, if we assume that in the Josephson
oscillation regime the nonlinear terms $Q_{1}^{\text{nl}}$ and $Q_{2}^{\text{%
nl}}$ have a rather small impact on the oscillation rate, we have the
following corrected inter-mode coupling strength
\begin{equation}
\tilde{\kappa}_{i}=\frac{(E_{i}^{0}+B_{12})C_{12}+\kappa }{C_{12}^{2}-1}\,.
\label{EffKappa}
\end{equation}%
So long as $C_{ij}$ is nonzero, this corrected coupling strength $\tilde{%
\kappa}_{i}$ can differ significantly from $\kappa $. In particular, the
kinetic energy in the transverse direction will make a positive contribution
to $E_{i}^{0}C_{12}$ but a negative contribution to $\kappa $. As a result,
these two terms will largely cancel each other and hence the kinetic energy
in the transverse direction will not directly enter into $\tilde{\kappa}_{i}$%
. For example, for the dipole orientation $\varphi =\pi /2$, using the same
parameters as in our previous effective two-mode model, we obtain $\tilde{%
\kappa}_{1}\approx $ $\tilde{\kappa}_{2}\approx -2.5$. One can then
expect that the predicted Josephson oscillation rate will be about
one order of magnitude closer to our numerical value. Further taking
into account the nonlinear terms $Q_{i}^{\text{nl}}$, one can expect
the magnitude of the inter-mode coupling to be further reduced and
hence better agreement with simulation results can be obtained. For
example, $Q_{1}^{\text{nl}}$ contributes a term $-B_{12}|\psi
_{2}|^{2}C_{12}/(C_{12}^{2}-1)$ to the
inter-mode coupling strength. This contribution will partially cancel the $%
B_{12}$ term in $\tilde{\kappa}_{1}$ defined above. All these observations
make it clear that the zero-overlap approximation is the main reason why our
effective two-mode picture cannot be used to predict the rate of Josephson
oscillations.

Consistent with our simulation result that the Josephson oscillation rate
depends strongly on the trapping geometry, we also find that if we slightly
change the width of the fitting mode wavefunctions $\Phi_{1,2}(\mathbf{r})$
and hence the value of $C_{12}$, then the value of $\tilde{\kappa}$ may also
change significantly. This implies that if we take Eqs. (\ref{ITM(a)}) and (%
\ref{ITM(b)}) as an improved two-mode theory, then it is possible to refit
the mode wavefunctions and obtain the right Josephson oscillation rate. This
approach is however not appealing to us, because by working with Eqs. (\ref%
{ITM(a)}) and (\ref{ITM(b)}) we lose the simplicity of a two-mode picture.

So why our two-mode picture with the zero-overlap approximation can still
give the correct transition point between the QST regime and the Josephson
oscillation regime? The main reason lies in that our two-mode treatment
already self-consistently used much information from the simulation results.
In particular, in the QST regime, $\Lambda_{c}$ defined in Eq. (15) is
dominated by $\Delta E$ and hence scales with $\kappa^{-1}$, the same
scaling behavior as $\Lambda$. As such, the actual magnitude of $\kappa$
will not affect the ratio of $\Lambda/\Lambda_c$, the key index for a
two-mode theory. In the Josephson oscillation regime, $\Lambda_c \sim 1 $
because $\Delta E \sim 0$, and $\Lambda << \Lambda_c$ because of the large
magnitude of $\kappa$. This is also consistent with the condition for the
Josephson oscillation regime in the two-mode theory.


\section{Concluding Remarks}

To conclude, using three-dimensional numerical simulations, we have
investigated the structures of the ground state and the quantum
self-trapping phenomenon of a purely dipolar Bose-Einstein condensate
trapped in a double-well potential. The anisotropic nature of the
dipole-dipole interaction is seen to impact dramatically on the transition
between two dynamical regimes, namely, the Josephson oscillation regime and
the quantum self-trapping regime. This finding is the key result of this
work.

To gain useful insights we constructed a simple effective two-mode model to
understand the transition from the self-trapping regime to the Josephson
oscillation regime. Interestingly, though our two-mode picture is based on a
rough zero-overlap approximation, the transition point obtained from our
two-mode picture is in excellent agreement with our numerical simulations.

Somewhat expected, when it comes to the rate of the Josephson oscillations,
our effective two-mode picture is no longer valid. We traced its failure to
the zero-overlap approximation and discussed how we might be able to improve
our effective two-mode treatment and find a better agreement with simulation
results by lifting the zero-overlap approximation. However, by lifting the
zero-overlap approximation, the simplicity of a two-mode picture in
understanding the quantum self-trapping is lost.

There are now keen interests in the dynamics of BEC's in optical lattices.
In particular, the quantum self-trapping effect is known to play an
important role in determining whether or not a condensate can spread out in
an optical lattice \cite{RosenkranzPRA2008}. The anisotropic nature of the
quantum self-trapping of a dipolar BEC suggests that tuning the dipole
orientation can lead to the control of the self-trapping effect in optical
lattices and hence the control of transport properties of dipolar BEC's in
periodic potentials. We hope our work here may stimulate experimental
efforts along this line.

We thank Ryan Wilson for pointing out a missing factor of two in Eq.
(4) in an earlier version of the manuscript. This work is supported
in part by the Academic Research Fund (WBS grant No.
158-000-002-112) (BX, WB and BL), National University of Singapore,
by the start-up fund (WBS grant No. R-144-050-193-101/133) as well
as the ``YIA" fund (WBS grant No. R-144-000-195-123) (JG), National
Univ. of Singapore, and by the NSF of US and the Welch Foundation
(Grant No. C-1669) (HP).

\appendix

\section{FOURIER TRANSFORM OF DIPOLE-DIPOLE INTERACTION POTENTIAL}

In this appendix we briefly outline how to use Fourier transformations to
calculate the dipole-dipole interaction potential in our three-dimensional
simulations. Assume that all the magnetic dipoles are aligned along an
external magnetic field $\mathbf{B}(t)$, with
\begin{equation}
\mathbf{B}\left( t\right) =B\left[ \hat{z}\cos \varphi +\sin \varphi \left(
\hat{x}\cos \alpha +\hat{y}\sin \alpha \right) \right] ,  \label{A1}
\end{equation}%
where $\hat{x},\hat{y},\hat{z}$ are units vectors in a cartesian coordinate
system. Then the dipole-dipole interaction energy becomes
\begin{eqnarray}
U_{dd}(\mathbf{r}) &=&d^{2}\frac{r^{2}-3\left[ z\cos \varphi +\sin \varphi
\left( x\cos \alpha +y\sin \alpha \right) \right] ^{2}}{r^{5}}  \notag \\
&=&-4\sqrt{\pi }\frac{d^{2}\sin ^{2}\varphi \cos ^{2}\alpha }{r^{3}}(Y_{00}-%
\sqrt{\frac{1}{5}}Y_{20})  \notag \\
&-&i\sqrt{\frac{6\pi }{5}}\frac{d^{2}\sin ^{2}\varphi \sin (2\alpha )}{r^{3}}%
\left( Y_{2-2}-Y_{22}\right)   \notag \\
&-&\sqrt{\frac{6\pi }{5}}\frac{d^{2}\sin 2\varphi \cos (\alpha )}{r^{3}}%
\left( Y_{2-1}-Y_{21}\right)   \notag \\
&-&i\sqrt{\frac{6\pi }{5}}\frac{d^{2}\sin 2\varphi \sin (\alpha )}{r^{3}}%
\left( Y_{2-1}+Y_{21}\right)   \notag \\
&&-4\sqrt{\frac{\pi }{5}}\frac{d^{2}\cos ^{2}\varphi }{r^{3}}Y_{20}+\frac{%
d^{2}\sin ^{2}\varphi }{r^{3}},  \label{A2}
\end{eqnarray}%
where $Y_{lm}$ is the standard spherical harmonics. Consider then the
Fourier transform of $U(\mathbf{r})$,
\begin{equation}
\tilde{U}_{dd}\left( \mathbf{k}\right) \equiv \int U_{dd}\left( \mathbf{r}%
\right) e^{i\mathbf{k\cdot r}}d^{3}\mathbf{r}.  \label{A2-1}
\end{equation}%
To evaluate this Fourier transform we first use the following identity%
\begin{equation}
e^{i\mathbf{k\cdot r}}=4\pi \sum\limits_{l=0}^{\infty
}\sum\limits_{m=-l}^{l}i^{l}Y_{lm}^{\ast }\left( \theta _{\mathbf{k}},\phi _{%
\mathbf{k}}\right) j_{l}\left( kr\right) Y_{lm}\left( \theta ,\phi \right) ,
\label{A3}
\end{equation}%
where $j_{l}\left( x\right) $ is the spherical Bessel function, and $\theta
_{\mathbf{k}}$, $\phi _{\mathbf{k}}$ are two spherical angles that define
the direction of the $\mathbf{k}$ vector. Upon integrations over the radial
coordinate $r$ we finally have
\begin{eqnarray}
\tilde{U}_{dd}\left( \mathbf{k}\right)  &=&\pi d^{2}\left[ 2\sin ^{2}\varphi
\sin ^{2}\theta _{\mathbf{k}}\cos \left( 2\phi _{\mathbf{k}}+2\alpha \right)
+\right.   \notag \\
&& 2\sin 2\varphi \sin 2\theta _{\mathbf{k}}\cos \left( \phi _{\mathbf{k}%
}+\alpha \right)   \notag \\
&&\left. +\left( 4/3-2\sin ^{2}\varphi \right) \left( 3\cos ^{2}\theta _{%
\mathbf{k}}-1\right) \right] .  \label{A4}
\end{eqnarray}%
For our purpose here for which the external magnetic field is
time-independent, we set $\alpha =0$ and hence obtain Eq. (\ref{FDDICONV}).

\end{document}